\documentstyle[11pt,newpasp,twoside,psfig]{article}
\newcommand{\et}{\sl et al \rm}

\begin{document}

\title{The Ghost in the Machine}
\author{Andy Lawrence}
\affil{Institute for Astronomy, University of Edinburgh, Royal Observatory, 
Blackford Hill, Edinburgh EH9 3HJ, Scotland, UK}

\setcounter{page}{111}
\index{Lawrence, A.}

\begin{abstract}
I examine simple tests for the presence of accretion disks in AGN - changes 
of 
surface brightness with viewing angle, changes of colour temperature with 
luminosity, and behaviour during variability. AGN observations pass the 
first two 
tests but fail the third, unless there is some previously unobserved source 
of 
heating - the ``ghost in the machine''. 
\end{abstract}

\section{Introduction}

We would like to know if AGN really have accretion disks. Predictions of the 
spectral energy distribution (SED) are hard to make, and model 
dependent, and so are not a robust test. Instead, I consider effects that 
test the generic idea of blackbody emission at a range of temperatures from 
a flat 
surface. 

\section{Behaviour of surface brightness with viewing angle}

From a tilted flat surface, we should see a standard $\cos\theta$ dimming 
plus 
whatever atmospheric limb darkening effects might apply. For superluminal 
sources 
we can derive a fairly accurate 
viewing 
angle to the jet. For a sample of 21 such sources Rokaki \et (2002; see also 
these 
proceedings) have 
measured the $H\alpha$ equivalent width (EW), 
which gives the relative continuum brightness if the face on EW is always 
the same. 
The results show two separate effects. At small angles EW 
has a steep $\theta$ dependence, presumably due to relativistic jet beaming. 
At 
larger 
angles, EW declines more gently with $\theta$, and the simplest model, with 
$\cos\theta$ 
dimming and standard thin disk limb darkening, fits extremely well. The data 
are not 
good enough to distinguish rival disk models, but the general idea of a flat 
tilted 
emitter is strongly endorsed.

\section{Behaviour of colour temperature with luminosity}

The SEDs of luminous quasars peak in the UV, roughly as expected for 
blackbody 
emission from around a billion solar mass black hole. (Actually, observed 
SEDs are a 
little cooler than expected..). The spectral index curves slowly from 
$\alpha_{opt}=0$ 
in the optical, through $\alpha_{uv}=1$ near the peak, and finally 
steepening to 
$\alpha_{fuv}=1.8$ in the far-UV and onwards through the soft X-ray excess. 
(Elvis \et 1994). However lower luminosity Seyfert galaxies 
often have $\alpha_{opt}\sim 1.0$, and dwarf AGN have $\alpha_{opt}\sim 2$ 
(e.g. Ho 
Filippenko and Sargent 1996; Ho, these proceedings). In fact $\alpha_{opt}$ 
and 
$\alpha_{uv}$ change slowly and systematically with luminosity (Mushotzky 
and Wandel 
1989; Zheng and Malkan 1993; Puchnarewicz \et 1996; see Fig. 1). This change 
is 
consistent 
with the Big Blue Bump (BBB) shifting to lower frequency with lowering 
luminosity, 
from blue bump to green bump to red bump, in a manner reminiscent of a 
cooling 
blackbody.

Of course accretion disks are not single blackbodies, but the sum of 
blackbodies. 
One can show that if an SED $L(\nu)$ is made of the sum of blackbody 
emitting areas, 
and that each sub-area changes temperature by the same factor $X=T/T_0$, 
then the 
new SED can be expressed in terms of the old one by the scaling formula 
$L(\nu)=X^3 
L_0(\nu/X)$ (Lawrence 2002 in preparation). One expects accretion discs to 
behave 
just this way when varying accretion rate or central heating rate, and 
roughly this 
way when varying black hole mass, as long as one considers emission coming 
well away 
from the inner boundary. As a result, if 
one has a template SED, then one can use the scaling formula to predict all 
other 
SEDs without having to know how the SED was made - i.e. whether it was from 
a thick 
disk, donut, flared disk, etc. The scaling simply tests the idea that the 
SED is 
made from the sum of blackbodies.

\vskip -0.2cm
\begin{figure}[h]
\centerline{
\psfig{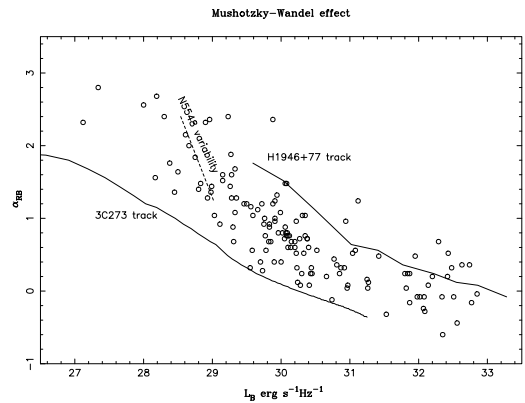}
}
\caption{\it Colour-luminosity effect in AGN. Here the ``R'' band is at 
$7500\AA$ 
and the ``B'' band at $4200\AA$. Data taken from Mushotzky and Wandel 1989. 
Model 
tracks as described in the text.}
\label{fig1}
\end{figure}

\vskip -0.2cm

I took two template SEDs - 3C273, from Kriss \et 1999, and H1946+786 from 
Kuhn \et 
1995 - and calculated the expected $\alpha_{opt}$ vs luminosity track. The 
result is 
shown in Fig. 1 and shows an impressive agreement remembering that this 
prediction 
has NO adjustable parameters. It starts to break down at the low end, but 
this could 
easily be due to boundary conditions, or the effect of reddening etc. The 
extreme 
ends of the diagram require the controlling parameter, e.g. accretion rate, 
to vary 
by something like a  factor of a million.

\section{Behaviour of colour temperature with variability}

Seyfert galaxy optical-UV variations show the BBB getting 
bluer when brighter. Do these variations also follow the sum-of-blackbodies 
scaling  law ? Fig. 1 indicates the low and high states of NGC 5548, from 
Peterson 
\et 
1991. Clearly the colour variations are far too strong. A single black body 
track is 
almost this 
steep, but the wide range of temperatures needed to explain the SED produces 
a far 
flatter track. A possible explanation is that the colour changes result from 
mixing 
a  variable blue component of approximately fixed shape with a static red 
component. 
Perhaps then the $\alpha - L$ locus we see in Fig. 1 
represents a kind of baseline stable state for AGN accretion discs heated by 
gravity/viscosity, with the thickness of the correlation caused by various 
degrees 
of mixing of a second variable component that has a different heating 
source. 

From the point of view of accretion disc models, or indeed any models where 
the 
energy is generated through an extended structure, there are two problems 
with 
Seyfert galaxy optical-UV variations. (i) The timescale - months - 
is far shorter than the viscous timescale expected  for stable accretion 
discs, 
which is hundreds of years. (ii) Variations occur simultaneously at 
different wavelengths, whereas emission at different wavelengths should 
arise at very
different radii. Fluctuations travelling at sound speed should then produce 
delays 
of order months. Krolik \et (1991) proposed that much of what 
we see is reprocessed radiation, so that fluctuations co-ordinate at light 
speed. 
That paper discussed optical fluctuations driven by UV, but most authors 
since have 
assumed that the variable UV itself is driven by variations in the observed 
X-ray 
source.
The reprocessing idea seems to have been dramatically 
confirmed by observations of NGC~7469, which shows that there {\em are} 
delays, but 
on a timescale of hours to days, and with delay varying as $\lambda^{4/3}$ 
exactly 
as expected for a simple centrally heated disc model (Collier \et 1998, 
Wanders \et 
1997). 

\section{The heating problem}

So accretion discs score well on the brightness-angle test, and on the 
luminosity-colour test, but fail to match observed variations, unless we 
drop the 
idea of local gravitational heating and instead have a second heating 
source, which 
dominates in the UV regions, and which communicates across the disk at light 
speed. 
For some years the second heating source has generally been identified with 
the 
observed X-ray source.
However, the X-ray heating model has two 
serious problems. First, simultaneous UV 
and X-ray observations of N7469 have not shown the expected correlation, 
with the 
X-ray emission if anything {\em trailing} the UV (Nandra \et 1998). Second, 
the 
X-ray 
luminosity is substantially less than the UV luminosity - by a factor of a 
few in 
low luminosity Seyferts, and by a factor of twenty in quasars. We arrive 
then at the 
strange conclusion that AGN accretion disks seem to 
be externally heated, but by an agent that we do not see. Furthermore this 
mystery 
agent must carry most of the primary energy in AGN. This then is the {\em 
ghost in 
the machine}.

What else could heat the observed variable UV source ? If we require 
external 
heating at light speed, there would be seem two possibilities - an unseen 
radiation 
source, or particles. The most obvious radiation source is EUV emission from 
the 
very inner disk. In observed terms there is just enough energy - 
essentially we are talking about the right hand side of the BBB heating the 
left 
hand side. However in flat discs, and even in flared discs, it is hard for 
the outer 
disc to see enough of the inner disc. (Gravitational bending helps, but 
probably not 
enough). Particles are an interesting and under-explored possibility. 	We 
want such 
particles to deposit their energy in the disc as heat before radiating, so 
protons 
are a much better bet than electrons. If they are not to cool on local 
electrons 
before hitting the disk, then they are probably high energy protons. One 
such 
possibility is virialised protons at $T\sim 10^{12}$K from a two-temperature 
inner 
region, if a thin cold disc can co-exist with it. This possibility has been 
considered by Spruit and Hardt (2000), although they assume that such 
protons heat 
an X-ray corona, which in turn heats the disc, rather than heating the cold 
disc 
itself. A second source of high energy protons could be from a standing 
shock near 
the black hole (e.g. Kazanas and Ellison 1986). It could well be that much 
of the 
accretion energy is generated in the ``plunging region'' inside 3$R_{Sch}$, 
with
most of the mass disappearing, but a large fraction of the energy being 
channelled 
into ultrarelativistic particles that escape. As these will follow field 
lines it is 
likely that most of these particles will end up in the disc. However they 
may lose 
most 
of their energy on the way, or at the skin of the disc, producing more 
gamma-rays 
than we observe. Finally, bombardment by such particles may significantly 
alter 
abundances in the disc. Skibo (1997) has in fact already argued that 
features in the 
wings of the X-ray Fe line can be attributed to emission from just such 
enhanced 
abundances produced by cosmic rays.
 
The above discussion assumes that the disc is {\em externally} heated, 
persuaded by 
the consistency of the time-delay versus wavelength result with the expected 
light-travel time. However the near-simultaneity of variations more 
generally drives 
us to drop the idea of distributed local heating. Another possibility then 
is that 
much of the energy is generated near the black hole in some unknown form, 
and 
transmitted through the {\em interior} of the disc, perhaps by fast shocks, 
emerging 
at a larger radius as (approximately) nice simple black-body radiation. This 
makes 
an AGN seem more like a star, but with a nastier geometry. Probably in fact 
we need 
both local and central energy generation.

\end{document}